\documentclass{emulateapj}
\usepackage{amsmath}

\journalinfo{Accepted for publication in  AJ, Aug 31, 2006}

\shorttitle{Taurus Quest/DBSP}
\shortauthors{}

\begin{document}

\title{A Distributed Population of Low Mass Pre-Main Sequence Stars near the 
Taurus Molecular Clouds}

\author{Catherine L. Slesnick\altaffilmark{1},  John M. 
Carpenter\altaffilmark{1},   Lynne
A. Hillenbrand\altaffilmark{1}, \& Eric E. Mamajek\altaffilmark{2}}

\altaffiltext{1}{Dept.\ of Astronomy, MS105-24, California Institute of
Technology, Pasadena, CA 91125}

\altaffiltext{2}{Harvard-Smithsonian Center for Astrophysics, 60 Garden St.,
Cambridge, MA 02138}

\email{cls@astro.caltech.edu}

\begin{abstract}

We present a drift scan survey covering a $\sim$5$^o$ $\times$ 50$^o$ region toward the
Southern portion of the Taurus-Auriga molecular cloud.   
Data taken in the $B,R,I$ filters with the Quest-2 camera on the Palomar 48-inch telescope
were combined with 2MASS near-infrared
photometry to select candidate young stars.
Follow-up optical spectroscopy of 190 candidates led to identification of 42 new low mass  
pre-main sequence stars with spectral types M4-M8,
of which approximately half exhibit surface gravity signatures similar to known Taurus stars while the other half 
exhibit surface gravity signatures similar to members of the somewhat older 
Upper Sco, TW Hya and Beta Pic
associations. 
The pre-main sequence stars are spread over $\sim$35$^o$,
and many are located well outside of previously explored regions.  
From assessment of the spatial and proper motion distributions, 
we argue that 
the new 
pre-main sequence stars  
identified far from the clouds
cannot have originated from the vicinity of the 
1--2 Myr-old subclusters which contain the bulk of the identified Taurus members, 
but instead represent a newly-identified area of recent star-formation near the clouds.

\end{abstract}

\keywords{open clusters and associations: individual (Taurus) -- stars: low mass, brown dwarfs -- stars: pre-main sequence}

\section{Introduction}

The Taurus-Auriga molecular cloud complex has, for decades, been considered the 
quintessential
example of low density, isolated star formation.  This fact along with the proximity
of Taurus 
(140 pc) and its position
in the northern hemisphere has caused Taurus to be one of the most often studied 
star-forming regions.
Several authors (e.g., Brice\~{n}o et al. 1999, 
Luhman 2000, Brice\~{n}o et al. 2002, Luhman et al. 2003, Luhman 2004, Guieu et 
al. 2006) have used optical/near-infrared/x-ray imaging 
to identify young star candidates within the clouds.  
Follow-up optical spectroscopy of photometrically-selected candidates
can distinguish 
members of Taurus 
from foreground or background field interlopers.
These studies found the Taurus population is clustered into several loose aggregates (Gomez 
et al. 1993), and is predominantly very young ($\sim$ 1--2 Myr; e.g., 
Brice\~{n}o
et al. 1998, Hartmann 2000).  
Thus far, no conclusive evidence has been established for a widespread population of older stars
within or near the cloud (Brice\~{n}o et al. 1999).

The phenomenon of short timescale (1--2 Myr) clustered
star-formation is not unique to Taurus but has been found in almost all other nearby young associations
(e.g., Carpenter 2000, Palla \& Stahler 2000).
The large numbers of very young stars and apparent lack of more evolved (5--10 Myr-old)
objects in star forming regions
contrasts with 
ages of a few tens of megayears (e.g., Blitz \& Shu 1980) inferred for molecular
clouds.
Either star-formation takes place for only a small fraction of the
cloud lifetime, or molecular clouds themselves live only a few megayears (e.g., Hartmann et al. 2001).
This problem has been discussed in the literature 
for almost three decades, and is commonly known as the `Post T-Tauri Star 
Problem'
(Herbig 1978). 

An alternative explanation for the apparent lack of older stars in molecular 
clouds is that such objects have been missed
in previous surveys.
Limitations in telescope time and instrument fields of view have constrained 
most previous deep
imaging surveys in Taurus (e.g., Brice\~{n}o et al. 1999, Luhman 2000) 
to small areas focused on subclusters (each $\sim$1 pc wide) 
that contain most of the known 
young members.
Assuming a mean
velocity dispersion of $\sim$2 km/s (Hartmann et al. 1986), a 
putative population of 5--10 Myr old stars in Taurus could travel
up to $\sim$20 pc (8 deg) away from its birth site.

Several studies (e.g., Neuh\"{a}user 
et al. 1997, Wichmann et al. 1996) have attempted to use the ROSAT All Sky Survey (RASS)
with spectroscopic follow-up to identify G--mid K type 
post T-Tauri Stars (stars with ages $\sim$3--10 Myr; PTTSs) far from the current Taurus members.
These observations revealed a distributed population of lithium-rich stars (indicating that they 
are younger than $\sim$100 Myr) 
that are widely dispersed across the cloud and beyond.
However, because both the decay of x-ray emission and the depletion of lithium
occur slowly for G--type stars with ages $<$100 Myr, 
these data alone cannot discriminate whether the RASS-selected stars
represent a 1 Myr-old or 10 Myr-old Taurus population, or a 100 Myr-old population that originated 
elsewhere.
Consequently,  
the origin of these stars and their relation to Taurus is still controversial (e.g., 
Brice\~{n}o et al. 1997).
The largest optical/near-infrared imaging survey to date that searched for  
Taurus members  
is that of Luhman (2006) who used a combination of USNO and 2MASS 
magnitudes to find young brown
dwarfs within a 15$\times$15 square degree region centered on the known 1 Myr-old subclusters.
The survey was aimed specifically at identifying young brown dwarfs
with colors and magnitudes similar to known $\sim$1 Myr-old substellar Taurus members and
was not targeted at finding older objects or comparably-aged low mass stars.

We have completed a new optical $U,B,R,I$ imaging survey of $\sim$250 deg$^2$ near 
the Taurus molecular
clouds. 
The specific survey area 
was chosen to include both well-known subclusters of stars and 
regions
beyond previously studied parts of the cloud.  
The current work is a companion survey to
a photometric and spectroscopic study we have carried out in the Upper Scorpius 
star-forming region (Slesnick, Carpenter, \& Hillenbrand 2006; hereafter SCH06) where we identified
43 new brown dwarfs and low mass stars. 
Our primary goal in Taurus is to search for and characterize a possible distributed population
of pre-main sequence (PMS) stars within and surrounding the clouds.
We combine the Taurus photometric data with 2MASS $J,H,K_S$ photometry 
to select candidate young stars (Section 2).  In Section 3
we present newly obtained spectral data for 190 of these candidates from which we 
identify a population
of 42 low mass PMS objects.
We assess the spatial and proper motion 
distributions
of the new population in Section 4 and discuss why these objects may have been missed
in previous surveys.
 
\section{Quest Imaging Observations}

\subsection{Photometric Monitoring and Data Processing}

$U,B,R,I$ observations were obtained with the Quest-2 Camera 
(Rabinowitz 
et al. 2003) on the 48-inch Samuel Oschin Schmidt Telescope at Palomar 
Observatory.  The Quest-2 camera is a large-area mosaic consisting of
112 CCDs arranged in four columns of 28 detectors.
The camera covers a 3.6$^o \times$4.6$^o$
field of view, and taking into account gaps between columns and chips, the instantaneous 
on-sky coverage is 9.4 deg$^2$.  Each of the four columns views
the sky through a separate filter (Johnson $U,B,R,$ or $I$ in this case, though
our data are calibrated to the Sloan photometric system; see below).
We operated the camera in drift scan mode where the final data
product is a strip of uniform width in declination and time-dependent length in RA 
that has been imaged in four filters.

One strip covering the RA range 40$^o\lesssim\alpha\lesssim$90$^o$
and spanning 4.6 deg in declination 
centered on $\delta$=22.5$^o$ was observed twice per night 
on the nights of 27-30 November 2003 and 4-5 December 2003 for a total of 12 scans of the
same portion of the sky.  This spatial area includes the young regions of L 1536 and L 1529,
as well as the Pleiades open cluster ($\alpha$=57$^o$, $\delta$=24$^o$).  
The CCDs are less sensitive in the $U$-band than anticipated and few sources 
were detected.
We therefore exclude the $U$ 
data from the remainder of this discussion.
Fourteen of the 84 $B,R,I$ CCDs have failed since installation due to bad 
connections or 
faulty chips, rendering our spatial coverage within the survey 
region non-uniform.

Data reduction, source extraction, and photometry were carried out using the 
Yale reduction pipeline as outlined in SCH06.
After bias subtraction, dark correction and flat fielding (see SCH06 for more details), 
the pipeline performs aperture photometry for all stars 
through an aperture of half-width 3.5 pixels.
We have matched detections within a 0.8'' radius from the 12 different scans of 
the survey region. 
For each source we averaged coordinates
from individual scans. The typical astrometric RMS deviation
about the mean is $\sim$0.13'' 
for stars detected in at least half of the scans.  The average offset between Quest-2 and 2MASS
coordinates (see Section 2.2) is +0.2'' in RA and +0.04'' in Dec.

To account for non-linearity and pixel-to-pixel variations within a chip,
and zero point and color variations between chips, we 
matched a subset of data ($\sim$600,000 stars) taken by the Quest-2 collaboration (Rabinowitz et al. 2003)
in a different part of the sky 
(-2.5$^o\lesssim\delta\lesssim2.5^o$, 120$^o\lesssim\alpha\lesssim240^o$) 
to the Sloan Digital Sky Survey (York et al. 2000).
For each 
of the Quest-2 CCDs we computed a conversion from Quest-2 to Sloan magnitudes in 
the
form of:

$r=a_R + b_R \times R_{Quest} + c_R \times Row + d_R \times (R-I)_{Quest}$

$i=a_I + b_I \times I_{Quest} + c_I \times Row + d_I \times (R-I)_{Quest}$

$b=a_B + b_B \times B_{Quest} + c_B \times Row + d_B \times (B-R)_{Quest}$,

\noindent where $a_X,b_X,c_X,d_X$ are constants and `$Row$' refers to a row of pixels (perpendicular
to the drift scan direction)
 on the CCD.
Only stars with both Sloan and Quest-2 photometric uncertainties $<$0.1 mag, and instrumental Quest-2 
magnitudes between 11.5
$<R,I<$ 19 and 14 $<B<$ 21 were used to derive the calibrations.  
Constants were computed in discrete bins of 
0.75  mag, 50 pixels and 0.2 mag in color and determined
in an iterative manner for each equation above 
until each constant changed by $<$0.0005 mag for at least 3 iterations.
We applied the derived calibrations to the Taurus scans by linearly 
interpolating between values
in each parameter.
  
The Taurus data were taken primarily under non-photometric conditions.  
However, because the data consists of repeated scans, self-calibration is possible.
To account for variable conditions during a night or from night to night, 
we applied a photometric offset as a function of RA to each scan. 
For a given CCD, the Yale reduction pipeline divides a drift scan into discrete frames
2048 pixels in length.  
Due to edge effects, along with the fact that some of the data were taken through thick clouds
($\geq$2 mag of extinction), most
sources were not detected in all 12 scans.  
The `best' (i.e., most photometric) scan,
taken on 
27 Nov 2003, was used to define 
the photometric reference system.
We selected a subset of $\sim$100,000 stars from the source catalog that were 
detected in at 
least 6 
of 12 scans (including the reference scan)
and had no neighbors within 5''.
For each such star we computed the difference between the reference 
magnitude 
and the magnitude measured in an individual scan.

For every chip and scan we created a catalog of photometric offsets by stepping
through in RA every 5 local calibrator stars (typically spanning$\sim$0.5 deg) 
and calculating a median offset value   
between an individual scan and the reference scan.  Typical values were 0.18 mag in $r$, 0.15 mag in $i$, or 0.23 
mag in $b$.
We applied these offsets to the entire data set as a function of RA, scan, and CCD 
by linearly interpolating between values in RA.  
Offsets as a function of declination were also computed.
However, upon examination, we found no systematic structure in the declination offsets 
and thus, only the RA offsets were necessary. 

The photometric  
precision can be assessed
from the repeatability of observations for individual 
stars.
For each source, we computed an uncertainty-weighted average and RMS deviation of individual calibrated
measurements.
The RMS deviation can be 
affected by many
factors, including photometric noise, uncertainties in the derived 
Quest-2--to--Sloan calibrations or 
scan-to-scan weather calibrations, and intrinsic photometric variability 
of the star.  Because the 112 CCDs within the mosaic are of
varying quality, defining `typical' RMS deviation is meaningful only on a chip-by-chip basis.
From analysis of RMS values as a function of magnitude for all CCDs,  
average RMS values for most chips range from 0.03 to 0.1 mag
for stars brighter than
$r,i\sim$20 and $b\sim$21, and are as high as 
0.2 mag for a few CCDs. Scatter increases toward
fainter magnitudes as expected (see Fig. 2 in SCH06).

The accuracy of the absolute photometry is harder to quantify.  While we 
have accounted for relative extinction due
to weather within our dataset, we are not able to account for 
zero point shifts between the Taurus drift scans and the scans used to derive the
Quest-2--to--Sloan calibrations.  Despite this fact, comparison of the 
average $r-i$ color for the calibrated Quest-2 photometry agrees to $\sim$0.01 mag
with the average $r-i$ color of the $\sim$600,000 stars from Sloan used to derive the calibrations.
Therefore, while we cannot claim the photometry is 
on a standard Sloan $b,r,i$ system, it should be fairly closely 
aligned with Sloan.  
Nevertheless, we select candidate PMS stars based on a combination of $relative$ 
Quest-2 optical colors and absolute 2MASS near-infrared
colors (see Section 2.2).
The final source catalog contains 
$\sim$2.2 million sources within the $\sim$250 deg$^2$ survey region with both $r$-- and $i$--band photometry. 

\subsection{Candidate Pre-Main Sequence Selection}

\begin{figure}
\begin{center}
\scalebox{0.55}{\includegraphics[angle=-90]{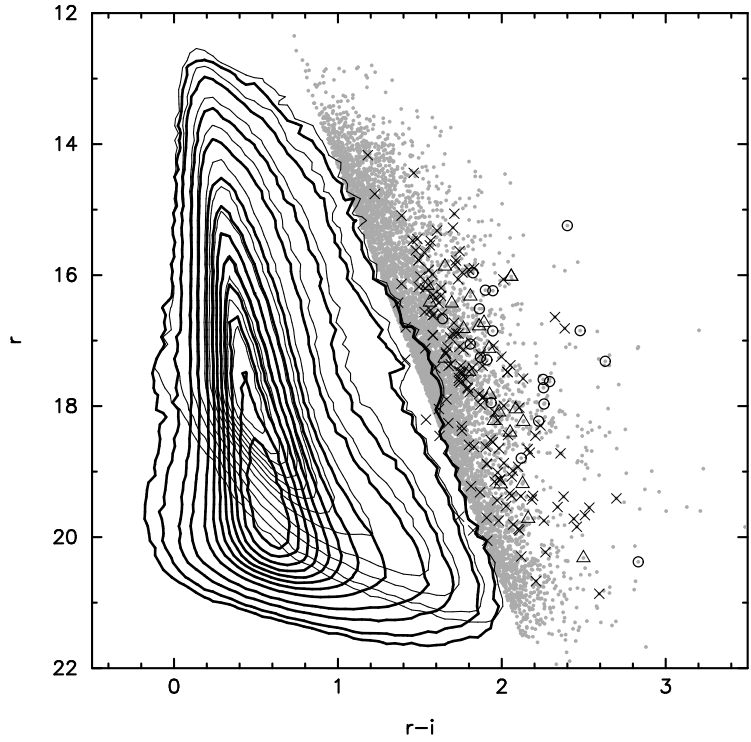}}
\end{center}
\caption{Optical color-magnitude diagram of all $\sim$2.2 million 
Quest-2 sources with $r$ and $i$ detections
(thick contours), and those with 2MASS 
counterparts (thin contours).
Contours represent data at 90\%--10\%, 5\%, 2\%, and 1\% of the peak level. 
Objects redward
of a linear approximation of the 1\% contour are shown as discrete grey points.
Objects for which we have spectral data are shown as large symbols: 
spectroscopically confirmed 
young stars are shown as circles, intermediate-age stars as triangles 
and dwarf stars as X's. }
\label{fig:cmd}
\end{figure}

Our goal is to use the photometry to isolate 
PMS stars both within and surrounding 
the Taurus clouds
from the field star population which dominates the source catalog.
Nearby young stars still undergoing contraction are systematically more luminous than their main-sequence counterparts, and will therefore 
occupy a sequence in an optical color-magnitude diagram (CMD) 
that is correspondingly brighter than the sequence occupied by 
most of the field stars. 
Due to challenges in calibrating the photometric data to a standard system, we 
were reluctant to choose candidates based on the colors and magnitudes
that correspond to particular isochronal ages.
For the initial spectroscopic observations
we considered for candidate PMS stars all sources redward of a linear approximation of the 1\% 
data contour in an $r,r-i$ CMD (see Fig.~\ref{fig:cmd}). 

In addition to optical colors, the infrared 
colors and magnitudes of potential PMS candidates were also considered.
We matched the entire 
source catalog to 2MASS and excluded 
from further consideration $\sim$1.2 million sources which did not have a 2MASS 
counterpart.
In Fig.~\ref{fig:cmd},
optical CMDs for all sources, 
and for sources with a 2MASS counterpart
are shown.  
As discussed in SCH06, requiring a 2MASS detection biases the list of potential PMS candidates against faint blue sources 
but does not exclude
red objects bright enough to be observed spectroscopically at Palomar 
($r\lesssim$20.5).  
The position of each star 
on a
near-infrared color-color diagram was examined and any star with $J-H,H-K_S$ colors 
consistent with those of background
giants [$(J-H) > 0.6(H-K_S)+0.6$ or $(J-H) > 1.69(H-K_S)+0.29)$] was excluded. We 
additionally considered $r-K_S$ colors 
and adopted the selection criterion $r < 2.57(r-K_S-3)+13$ as outlined in SCH06.  
After 
all selection criteria were applied, the final candidate list 
contains $\approx$1800 stars for spectroscopic 
follow-up.

\begin{figure}
\begin{center}
\scalebox{0.55}{\includegraphics[angle=-90]{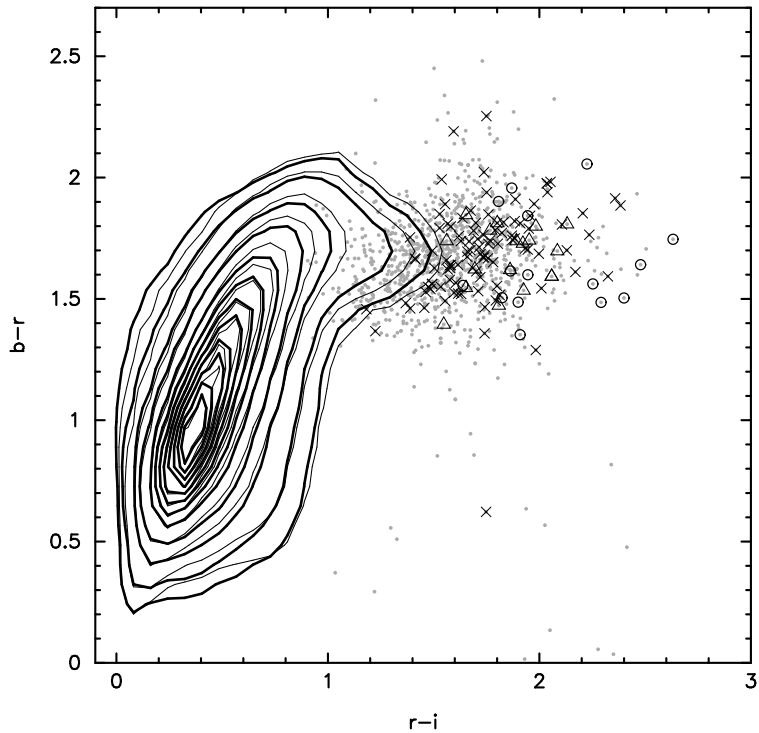}}
\end{center}
\caption{Optical color-color diagram of the $\sim$1 million Quest-2 sources with
$r,i$, and $b$ data (thick contours) and of those with 2MASS 
counterparts (thin contours)
represented at 90\%--10\%, 5\%, 2\%, and 1\% of the peak level. Objects
which meet all of the ($riJHK_S$) selection criteria to be candidate PMS stars 
and which have $b$ data
are shown as discrete grey points.  Large symbols are as in Fig.~\ref{fig:cmd}.}
\label{fig:ccd}
\end{figure}

Although $b$--band data was not used in candidate PMS selection,
$\approx$60\% of our candidates and 136 out of 190 objects 
observed spectroscopically (see Section 3)
have a $b$ detection.  Fig.~\ref{fig:ccd}
shows an $(r-i)$ vs. $(b-r)$ color-color diagram for all stars with $r,i,b$ detections, 
and for stars which additionally have 2MASS detections.  
Data with $b$--band detection
meeting the selection criteria 
outlined above are shown as discrete points.  In principle $b$--band (and $u$--band) photometry could be used
to select candidate young stars based on blue excess attributed to accretion (e.g.,
Rebull et al. 2000), though we have not implemented any such criteria in the present work.

\section{Optical Spectroscopy}
Moderate-resolution spectra of 190 candidate PMS objects (chosen from the $\sim$1800
candidates discussed in Section 2.2) were taken with the Double 
Spectrograph on the Palomar 200-inch
telescope on the nights of 2004 9-11 Dec and 2005 23-27 Nov.
Candidates were prioritized by $r-K_S$ color with redder candidates observed first,
and targets were 
selected primarily from the magnitude range 
15$ < r <$19.  The spatial distribution of observed stars is discussed in Section 4.1.
Data were taken with the red side of the spectrograph through a 2'' slit using a 
5500\AA$\;$
dichroic and a 316 lines mm$^{-1}$ grating blazed at 7500\AA$\;$.  This set-up 
produced wavelength coverage from
6300--8825\AA$\;$at a resolution of $R\sim$1250.  Typical exposure times were 
300--900 sec, and up to 1800 sec for the 
faintest targets ($r\sim$21).
Spectrophotometric standard stars (Massey et al. 1988) were observed throughout 
each night
for flux calibration.  All sources were processed, extracted and calibrated 
using standard IRAF
tasks.  

In addition to our program targets
we have also observed a range of spectral main sequence standards (K5-L3), giant
standards (K7-M9), and previously identified Taurus objects (K3-M7.25; 
Brice\~{n}o et al. 2002, Luhman 2004).
We observed several members of the Hyades ($\sim$650 Myr; Lebreton et al. 2001), 
Pleiades ($\sim$115 Myr; Basri, Marcy, \& Graham 1996), AB Dor ($\sim$75--150 
Myr; Luhman et al. 2005),
Beta Pic ($\sim$11 Myr; Ortega, et al. 2004) and TW Hya ($\sim$8 Myr; de la 
Reza et al. 2006) associations.
In the SCH06 survey 
we 
observed $>$50 members of 
the $\sim$5 Myr Upper Scorpius association.  Together, these observations
provide a large set of late spectral type (K--M) standards spanning a broad range of 
age (surface gravity).

\subsection{Spectral Analysis}
\begin{figure}
\scalebox{0.4}{\includegraphics{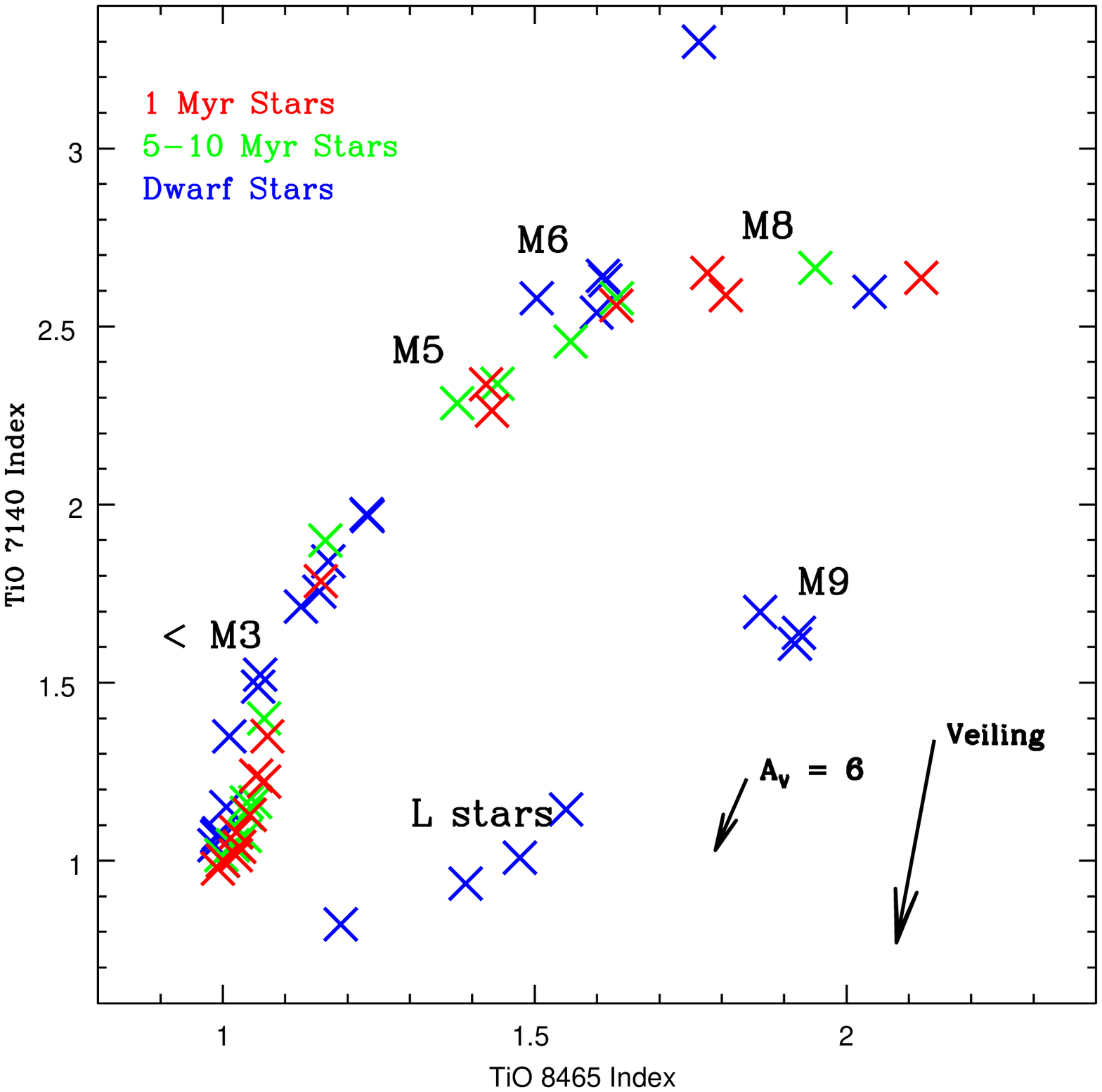}}
\hspace*{0.2in}
\scalebox{0.4}{\includegraphics{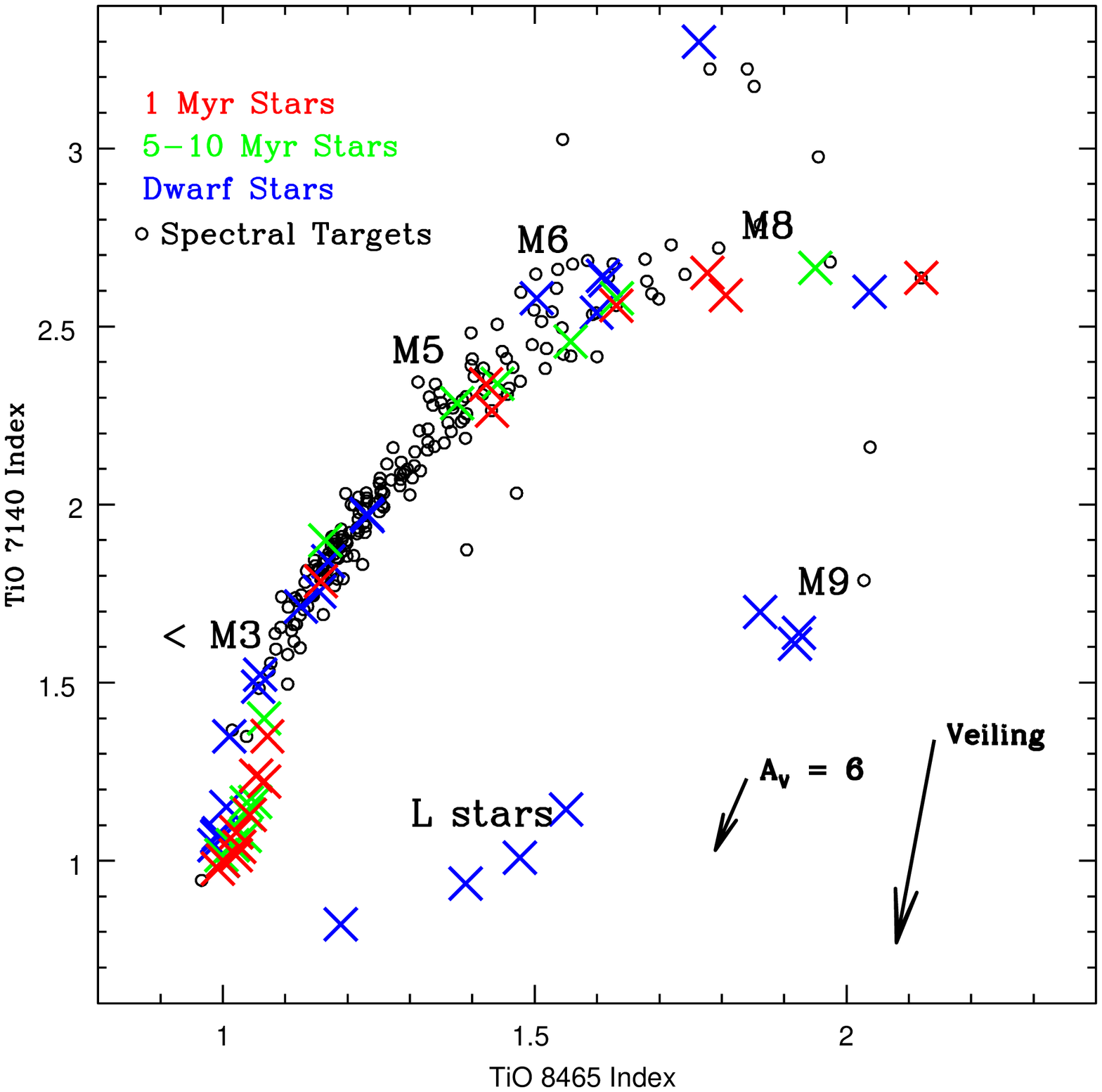}}
\caption{Temperature-sensitive TiO-7140 vs. TiO-8465 indices.  Blue X's 
represent measured indices for field dwarfs and
members of the Hyades ($\sim$700 Myr), Pleiades ($\sim$115 Myr) and AB Dor 
($\sim$75--150 Myr) associations.
Green X's show measured indices for intermediate-age spectral standards from 
Beta Pic ($\sim$11 Myr), TW Hya ($\sim$8 Myr),
and Upper Sco ($\sim$5 Myr).  Red X's show measured indices for young Taurus 
members ($\sim$1--2 Myr).  In the right 
panel, black symbols are measured indices for Quest-2 PMS candidates.
The effects of 
extinction and veiling are shown as vectors (see text). This diagram is useful for
classifying stars with spectral types M3-L3.}
\label{fig:temp}
\end{figure}

A detailed discussion of our classification methods is given
in SCH06, and thus only a summary is provided here.
We use the strength of titanium oxide (TiO) absorption which 
increases from mid-K through $\sim$M7 spectral types (see Fig. 8 of SCH06) as the primary spectral 
type diagnostic.
We have adopted from the literature 
two spectral
indices that measure the strength of TiO features: TiO-7140 
($F_{\lambda7035}/F_{\lambda7140}$
with bandwidths of 50\AA; Wilking et al. 2005) and TiO-8465 
($F_{\lambda8415}/F_{\lambda8465}$
with bandwidths of 20\AA; SCH06).
The left side of Fig.~\ref{fig:temp} shows a plot of TiO-8465 vs. TiO-7140 for 
observed spectral standards.
This diagram is 
useful for classifying stars 
with spectral types M3--L3.  No significant age dependence exists between 
measured TiO indices of older stars (ages $>$75 Myr; blue X's) 
and young (ages 1--2 Myr; red X's) or intermediate-age (ages 5--10 Myr; green X's)
PMS standard stars.

Measured 
indices for program sources are shown in the right side of Fig.~\ref{fig:temp}. 
The measurements predominantly follow the locus determined by the spectral 
standards.
Two outliers sit below the primary sequence of data points.  Both objects 
are confirmed to be
young stars (see below) with strong H$\alpha$ emission and we attribute their position 
in Fig.~\ref{fig:temp}
to a small amount of veiling present in their spectra (see Section 3.2).
Spectral types were first estimated from quantitative analysis of 
the measured TiO indices.  More weight was given to the TiO-8465 index, 
which is the  
less sensitive of the TiO indices to the effects of reddening and veiling (Section 3.2).  Final 
spectral types were
confirmed after visual inspection of each spectrum in comparison to spectral 
standards.

In addition to determining spectral types it was also necessary to determine 
which of the candidates
are bona fide PMS stars. 
Because young stars are still undergoing contraction to the main sequence, they 
have systematically lower surface gravity than older main sequence stars.  
We therefore use surface gravity 
to roughly define stellar age.  Several diagnostics of surface gravity exist 
in this wavelength regime
which can be assessed in low and moderate resolution spectra.  
The most prominent gravity-sensitive features in the Palomar spectra are K I 
(7665 and 7669 \AA) and Na I (8183 and 8195 \AA) absorption doublets.

\begin{figure}
\scalebox{0.4}{\includegraphics{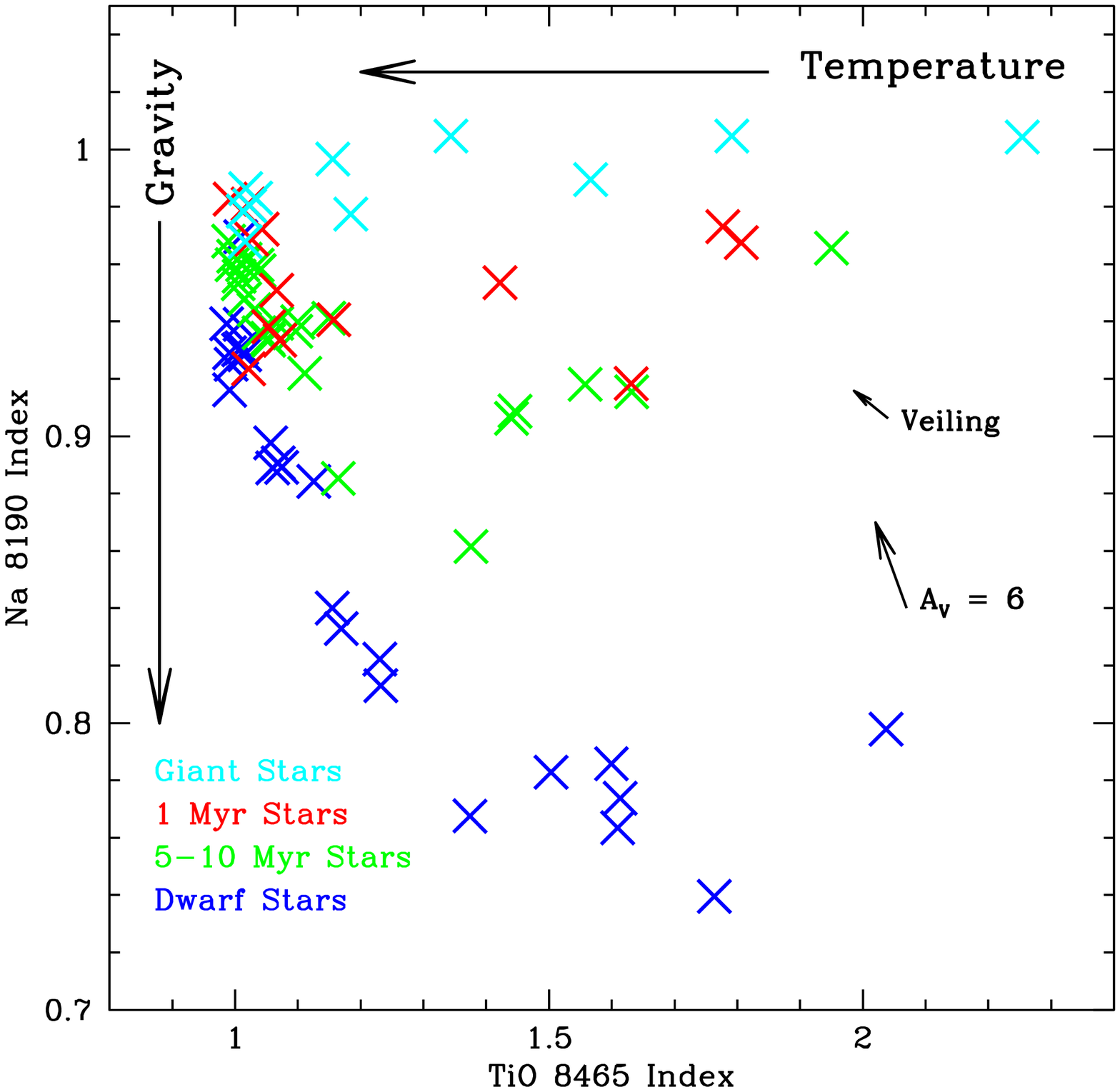}}
\hspace*{0.2in}
\scalebox{0.4}{\includegraphics{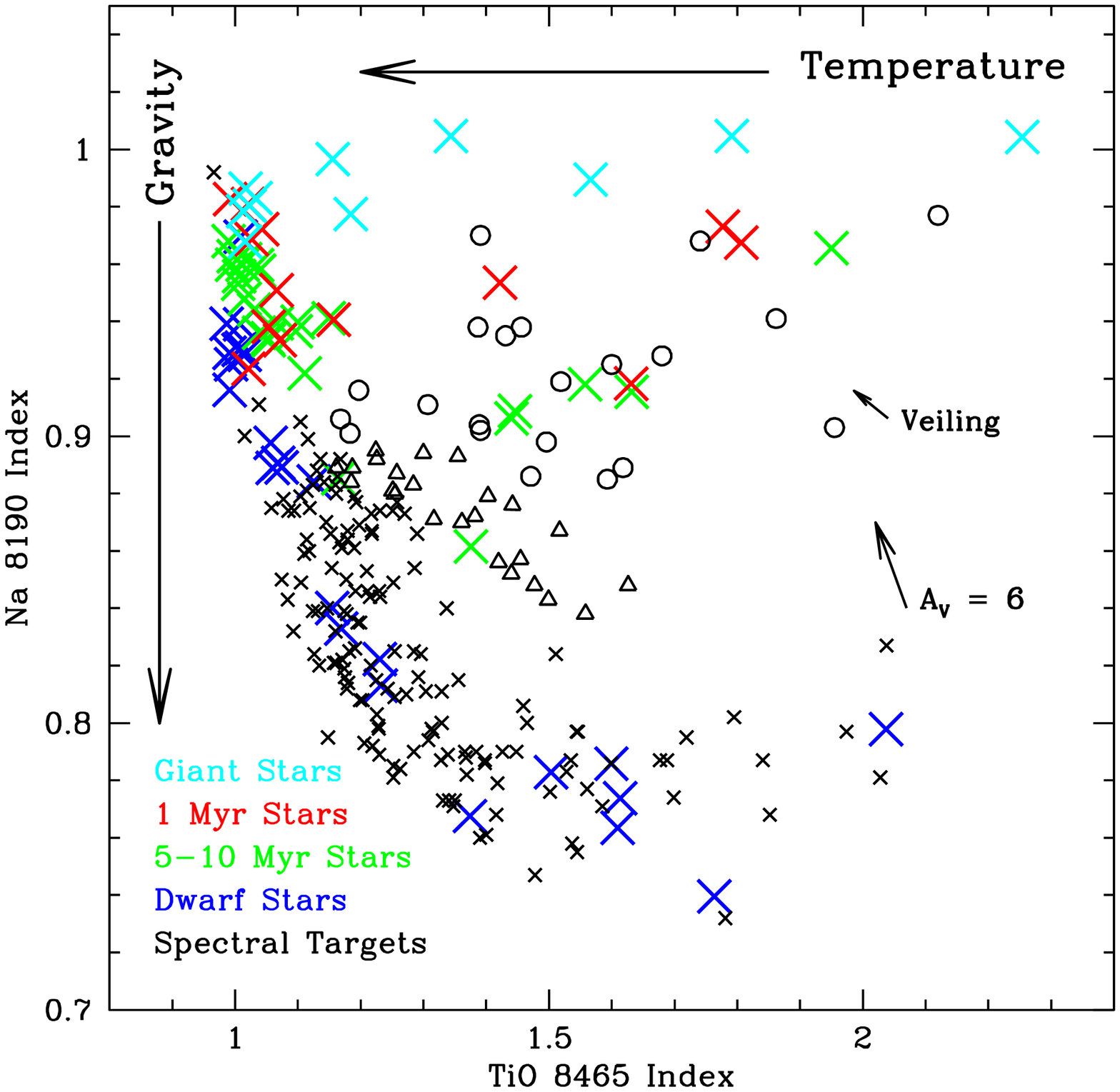}}
\caption{TiO-8465 index vs. the gravity 
sensitive Na-8189 index.  Symbols are as in Fig.~\ref{fig:temp} with the addition of
cyan X's which represent measured indices for giant standard stars.
Objects with higher surface gravity have more Na I absorption present in their spectra and thus,
a smaller Na-8189 index. In the right panel, black symbols representing program candidates
are divided into three groups based on inferred surface gravity:
circles represent objects with surface gravity lower (i.e., younger in age)
than the Upper Sco association, triangles indicate objects with intermediate
surface gravity comparable to the Upper Sco, TW Hya and Beta Pic associations,
and black X's indicate field stars.
As in Fig.~\ref{fig:temp}, the effects of veiling and extinction are shown
as vectors (see text). 
}  
\label{fig:grav}
\end{figure}

We assess surface gravity quantitatively by using  
the gravity-sensitive Na-8189 index ($F_{\lambda8189}/F_{\lambda8150}$
with bandwidths of 30\AA; SCH06) to measure the strength of the Na I doublet.
Fig.~\ref{fig:grav} shows 
a plot of the temperature-sensitive TiO-8465 index as a function of the gravity-sensitive 
Na-8189 index for all stars
earlier than M9.  On the left is shown measured indices for spectral standards.  
Colors are as in Fig.~\ref{fig:temp} 
with the addition of cyan X's for giant stars.  The right side 
of Fig.~\ref{fig:grav} shows measured indices for 
objects that are classified as having low (circles) or intermediate (triangles) gravity, 
and candidates spectroscopically determined to be field dwarfs (black X's; see below).  

One difficulty in interpreting Fig.~\ref{fig:grav} directly is that the Na-8189 index is 
contaminated by telluric features.  This point was not discussed in SCH06 because both the candidates
and Upper Sco spectral standards were taken at similar airmasses and thus this effect did not
influence the spectral classifications.
However, the Taurus spectral data were taken at systematically lower airmasses
than many of the spectral standards.
As a result, a program star with intermediate
gravity signatures observed at low airmass 
will have a lower Na-8189 index than an intermediate gravity standard 
of the same spectral type observed at high airmass.  
This effect causes a discrepancy between the positions of the green
X's and the triangles in Fig.~\ref{fig:grav}.  

\begin{figure}
\epsscale{0.8}
\plotone{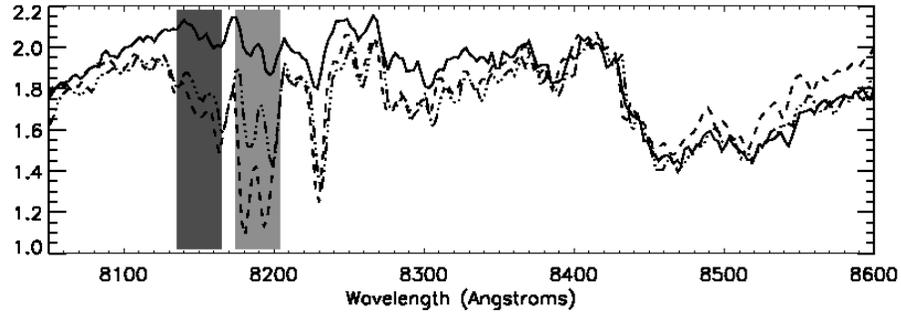}
\caption{Section of the optical spectrum highlighting the surface gravity sensitive Na I doublet
(8183 and 8195 \AA).
The dashed spectrum is an M5V star (GJ 866; Kirkpatrick et al. 1991), 
the dotted-dashed spectrum is an M5 Upper Sco member 
(USco CTIO 53; Ardila et al. 2000), and the solid spectrum is an M5 Taurus members 
(MHO 7; Brice\~{n}o et al. 1998).  Light and dark shaded 
regions show the respective locations of the Na I and the continuum bands used in constructing 
the Na-8189 index.  All spectra have been normalized at 8410 \AA$\;$near the temperature-sensitive
TiO (8465 \AA) molecular absorption band. 
Both GJ 866 and USco CTIO 53 were observed at high airmass and telluric absorption (8161-8282 \AA)
affects 
both the continuum band and Na I band, causing systematically high measurements of 
the Na-8189 index.  However, gravity signatures in the three spectra can be distinguished 
clearly through visual
inspection of the line strengths.  
}
\label{fig:gspeca}
\end{figure}

Fig.~\ref{fig:gspeca} shows a  
section of the spectra which highlights the Na I
(8183 and 8195 \AA) feature for dwarf, intermediate, and low gravity stars of the same spectral type.
Both GJ 866 and USco CTIO 53 (the dwarf and intermediate-gravity stars) 
were observed at high airmass whereas MHO 7 (the low gravity Taurus member) 
was observed at low airmass. 
Telluric absorption (8161--8282 \AA) seen in the spectra of GJ 866 and USco CTIO 53 
affects
both the continuum band and Na I band causing systematically high measurements of 
the Na-8189 index. 
However, the three spectra can be clearly distinguished through visual
inspection of the NaI line strengths.  Thus, we used the quantitative indices as a rough guide only and
all final gravity classification was done by eye. 
All objects with
surface gravity features weaker than those of the intermediate-age standards are considered 
by us to have low gravity; objects with gravity features similar to those exhibited by
the intermediate-age standards are considered to have intermediate gravity.  
Based on this classification scheme we  
identify 42 new PMS stars (see Table 1).  Of these, 19 exhibit low surface gravity features 
and the remaining 
23 exhibit intermediate surface gravity features.
A detailed discussion of both populations is given in 
Section 4.

\subsection{Extinction and Veiling}
Although most of the known low mass Taurus members have low extinction (A$_V 
\sim$1; e.g., Kenyon, Dobrzycka, \& Hartmann 1994),  
the quantitative indices used in classification are affected by 
interstellar 
 reddening
which we must account for in the classification process.
To assess any potential effect,
we artificially 
reddened all
spectral standards by A$_V$=6 mag, the maximum extinction within our survey region as inferred from 
large-beam dust maps 
(Schlegel et al. 1998).  We then re-measured all classification indices.  
Results of this experiment for an M5 star are shown as vectors on 
Figs.~\ref{fig:temp} and ~\ref{fig:grav}.
Reddening at this level is not sufficient to shift the indices by 
more than $\approx$0.5 subclasses
in temperature, and can be easily identified upon visual assessment of  
overall spectral slope.

We must also consider the effects of veiling, either due to excess emission 
from an accretion shock or due to 
thermal emission from dust grains in a circumstellar disk.  In SCH06 we 
explored  possible biases introduced by these processes
on our spectral classification by adding to the spectra hot or cold blackbody 
emission of constant temperature, or
continuum excess of constant flux.  SCH06 found that veiling due to
thermal emission from dust
grains in a disk does not significantly affect the spectral classification indices used here.
Veiling from a hot accretion shock,
however, does systematically decrease the TiO-7140 index.
The maximum shifts
produced from these experiments are shown as vectors in Figs.~\ref{fig:temp} and ~\ref{fig:grav}.  
We believe this effect to be the cause of measured indices for SCH J0429595+2433080 
and SCH J0518028+2327126 lying beneath
the primary sequence of points.  In addition to this evidence for veiling, 
both stars are confirmed to have low 
gravity and strong H$\alpha$ emission consistent with young,  
possibly accreting objects.  
SCH J0429595+2433080 also has CaII triplet emission.

\section{New Pre-Main Sequence Objects}

\begin{figure}
\epsscale{0.8}
\plotone{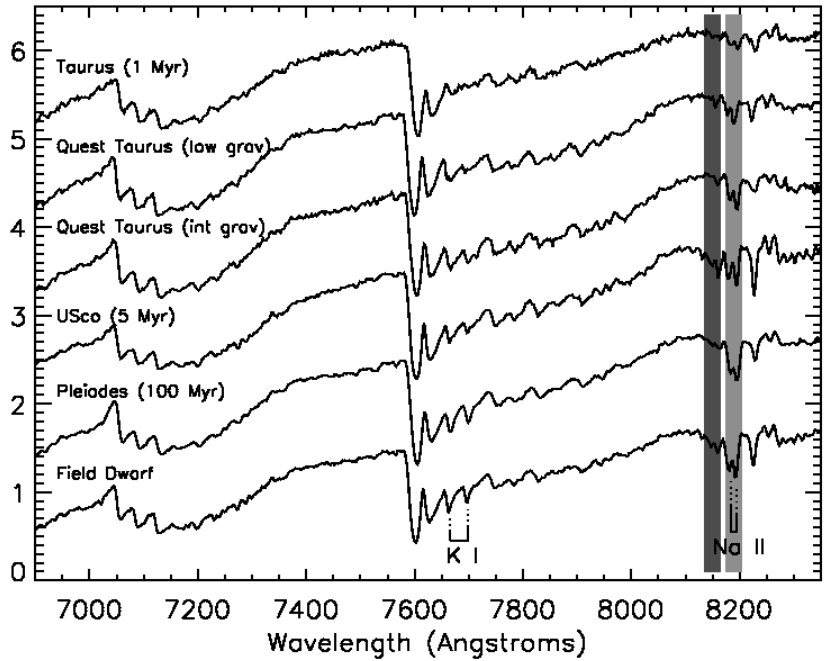}
\caption{Spectra of M4/M4.5 stars presented in order of decreasing surface 
gravity (bottom to top). 
Spectra shown are of a field dwarf, a 115 Myr Pleiades object, a 5 Myr Upper Sco 
member, a new Quest-2
Taurus candidate identified to have Upper Sco-type intermediate gravity, a new 
Quest-2
Taurus candidate identified to have low gravity, and a 1 Myr Taurus 
star.  
Surface gravity sensitive features include the KI doublet (7677 \AA) and the NaI doublet (8189 \AA).
}
\label{fig:gspec}
\end{figure}

From the spectroscopic data we identify 19 objects 
having spectral features which indicate lower surface gravity than members of Upper Sco ($\sim$5 Myr).
Most of the new low gravity objects have inferred gravities as low as 
those similarly inferred for known 
1--2 Myr Taurus members. 
Of these 19 objects, three were previously identified in the literature: SCH 
J04295950+2433080 (Guieu et al. 2006),
SCH J04311908+2335048 (Luhman 2006) and SCH J0416272+2053093 (Wichmann et al. 1996).  
We additionally identify 23 objects which have intermediate-strength surface gravity features 
consistent with those observed in 
Upper Sco, TW Hya and Beta Pic stars.
In Fig.~\ref{fig:gspec} we present spectra of  
M4/M4.5 stars
shown in order of decreasing surface gravity and 
decreasing Na I and K I absorption (bottom to top).   
All spectra of observed stars as old as AB Dor or the Pleiades ($\sim$100 Myr) appear 
identical to those of dwarf stars.  Therefore,
the intermediate-age population in Taurus is likely significantly younger than $\sim$100 
Myr, although the exact upper bound on the age of this population is unknown
due to lack of comparison stars with ages between 10 and 100 Myrs.
This interpretation may also apply to three objects identified by Luhman (2006) which were
found to have gravity intermediate between Taurus and dwarf stars and presumed in that study to be 
$\sim$100 Myr-old due to lack of comparison stars with ages between those of Taurus members and field dwarfs.  

Hereafter we refer to the low gravity objects as ``young'' and the
intermediate gravity objects as ``intermediate-age.''
In the color-magnitude and color-color diagrams (Fig.~\ref{fig:cmd} and
Fig.~\ref{fig:ccd}), candidates for which we have obtained spectral data are shown as large symbols. 
For both figures, objects confirmed to be young (including 
SCH J04295950+2433080,
SCH J04311908+2335048 and SCH J0416272+2053093) or intermediate-age PMS stars are 
distinguished from those determined to be older field objects.
Photometric and spectral data for new 
PMS stars is
given in Table 1. Appendix A contains magnitudes and spectral indice measurements for
PMS candidates spectroscopically determined to be field dwarfs.  

Understanding the relationship of the newly-identified PMS population to the known
Taurus members requires distances to the new PMS stars which we cannot determine based on the current 
data set.
If we assume they are located at the distance of Taurus, the derived ages from an HR diagram are
$\sim$1--10 Myr, and the intermediate-age stars tend to have systematically lower
derived luminosities than the young stars at a given spectral type.  
While these relative ages are consistent with those derived from 
surface gravity analysis, a range of distances could yield similar luminosity segregation results. 
In lieu of distance measurements, we constrain the origin of the PMS objects identified in this work
by assessing the projected spatial distribution and kinematics of 
the new young and intermediate-age stars 
in relation to the known Taurus population.

\subsection{Spatial Distribution}

\begin{figure}
\epsscale{1}
\plotone{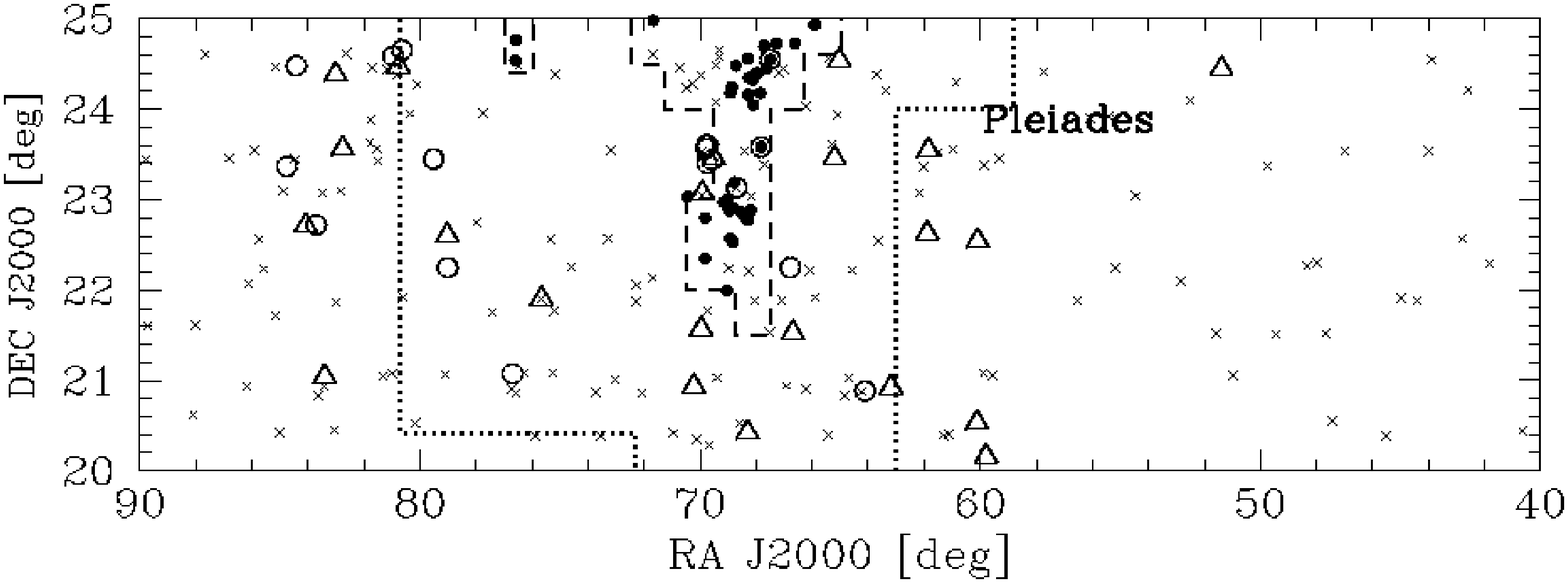}
\caption{Spatial area of the imaging survey shown with 
previously known low mass Taurus members (filled circles; see text for references).   
Dashed lines indicate the boundary of previous deep
CCD imaging surveys aimed at identifying new 1 Myr-old association members.  
Dotted lines indicate the boundary
a 5 Myr old star with  velocity 2 km/s could have traveled from any of the 
known subclusters.
Open circles and triangles represent 
new low and intermediate-gravity stars identified from this work.
Black X's show spectral candidates determined to be field dwarfs.
The location of the Pleiades cluster ($\alpha$=57 deg, $\delta$=24 deg) is 
indicated.}
\label{fig:spat2}
\end{figure}

The new young and intermediate-age stars are distributed 
throughout the survey region, and many are located 
well beyond regions previously explored for young 
PMS stars.
Fig.~\ref{fig:spat2} shows the location of spectroscopically observed Quest-2 candidates,
along with known low mass Taurus members from the literature 
(Brice\~{n}o et al. 2002, Guieu et al. 2006, Hartmann et al. 2002, Luhman et al. 
2003, Luhman 2004, Luhman 2006).
The region that has been previously studied for low mass 
stars 
using deep optical/near-infrared imaging with spectroscopic follow-up
(Brice\~{n}o et al. 2002, Guieu et al. 2006, Luhman 2000, Luhman et al. 2003, 
Luhman 2004) is indicated, as is the Pleiades association.
Based on comparison with CO maps of the region (Dame, Hartmann, \& Thaddeus 2001), while 
some objects do lie in projection near molecular gas, 
we see no systematic correlation between the spatial distribution of the new PMS stars 
and the CO emission.

\begin{figure}
\epsscale{0.8}
\plotone{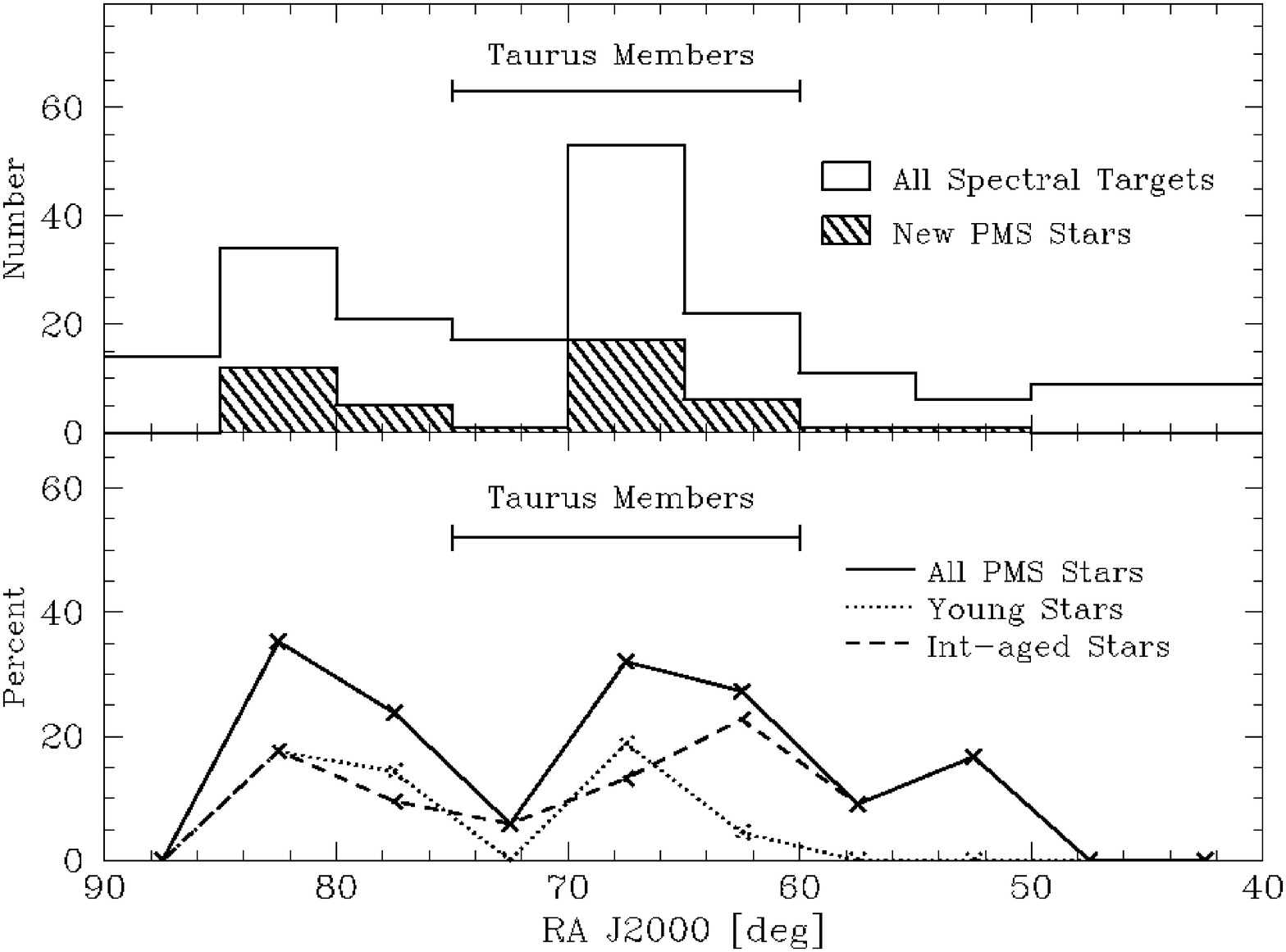}
\epsscale{0.8}
\plotone{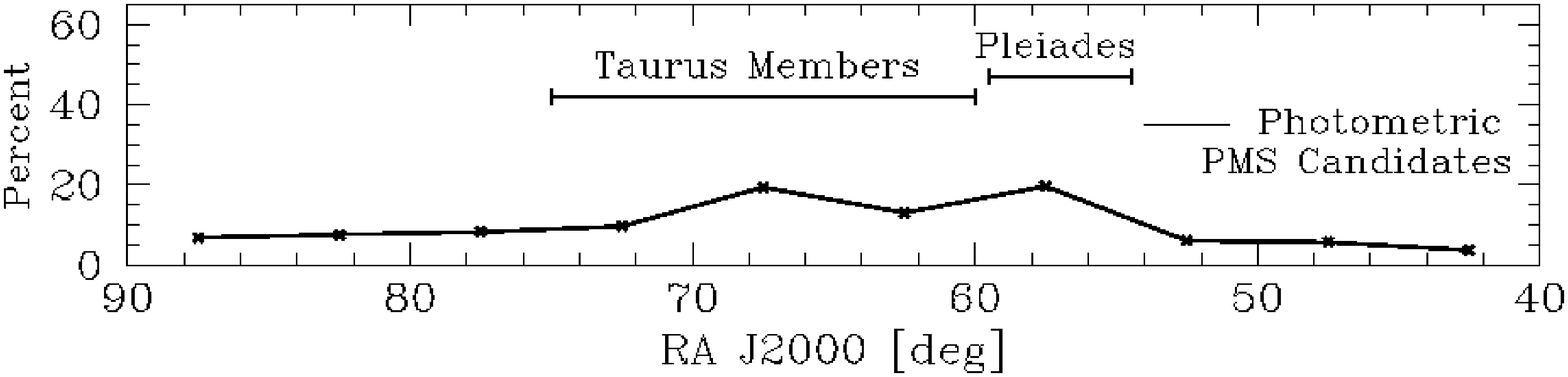}
\caption{Top panel shows, as a function of RA, histograms for the total number of stars observed 
spectroscopically (open histogram) and
those that were determined to be PMS stars (hatched histogram).
Middle panel indicates the percentage of spectroscopically observed objects 
classified as PMS stars (solid line), and
separately the percentage determined to be young (dotted line) or 
intermediate-age (dashed line) stars based on spectroscopic signatures of surface gravity.  
The RA range containing 98\% of known low mass Taurus members 
is shown.
Bottom panel shows the percentage of the $\sim$1800 photometric PMS candidates which fall at
a given RA.
}
\label{fig:sbin}
\end{figure}

To assess whether the new PMS stars are
associated with the known concentrations of Taurus 
members or whether they are more uniformly distributed,
we show in Fig.~\ref{fig:sbin} a histogram of the RA values for all 
sources with spectra presented here 
and for those sources confirmed  
as PMS objects.  In the middle panel we present the 
percentage of spectroscopically observed objects determined to be 
PMS 
stars.
These can be compared to the RA range that contains 98\% of the known low mass Taurus members
(Brice\~{n}o et al. 2002, Guieu et al. 2006, Hartmann et al. 2002, Luhman et al. 
2003, Luhman 2004, Luhman 2006).
The bottom panel shows the percentage of the 1800 candidates located
within a given RA range.
 
We note two spatial 
concentrations of new PMS stars: one near the known Taurus members at $\alpha \approx$68$^o$,
and a second 
in the Eastern portion of the cloud centered around $\alpha 
\approx 82^o$.
In Fig.~\ref{fig:spat2} we indicate the approximate boundary within which
a 5 Myr star with velocity 2 km/s relative to Taurus could have traveled from any of the 
known Taurus subclusters.
Some of the young objects newly identified here are
located well beyond this region.  Assuming an age of $<$5 Myr as derived
from surface gravity analysis, if these new young stars 
originated in the 
known star-forming subclusters they must, therefore, have arrived at their current positions at
relatively high velocities.  

An alternative, and perhaps more likely scenario is that we have identified previously 
unknown areas of recent star-formation outside of 
the current dense cloud complex.
To further quantify the new PMS population and assess this possibility,
we divide our PMS stars into 3 groups:  those that lie in the same RA range as
98\% of the known Taurus sample (60$^o \leq \alpha \leq 75^o$), those that are East 
($\alpha > 75^o$) and those that are West ($\alpha < 60^o$) of this RA range.
We find that of the 190 spectral candidates, 2/33 (6\%) are confirmed to be PMS stars in the Western region,
23/89 (26\%) in the central region, and 17/68 (25\%) in the Eastern region. 
Spectroscopic confirmation rates of 25\% in the East and only 6\% in the West are 
contrary to the isotropic distribution 
we would expect to observe if these stars had been dispersed from the central regions.
Using the two-tailed 
Fisher Exact test we compute a probability of $\sim$3\% that the observed Eastern and Western
distributions 
could have been drawn from the same population.
We can therefore conclude with 97\% confidence that the distributed PMS stars were not 
randomly dispersed from the known 1--2 Myr-old Taurus population.  
Instead,
we suggest that they likely represent a population
that is not associated with the currently visible areas of the dense Taurus-Auriga
molecular cloud complex.

\subsection{Proper Motions}

A primary goal of our large-area survey is to search for  
and characterize any
PMS stars that might exist far from the $\sim$1--2 Myr-old subclusters in Taurus.
Having identified several tens of such stars, we can use proper motion information to further study the 
characteristics of this
spectroscopically-selected sample.

\begin{figure}
\epsscale{0.8}
\plotone{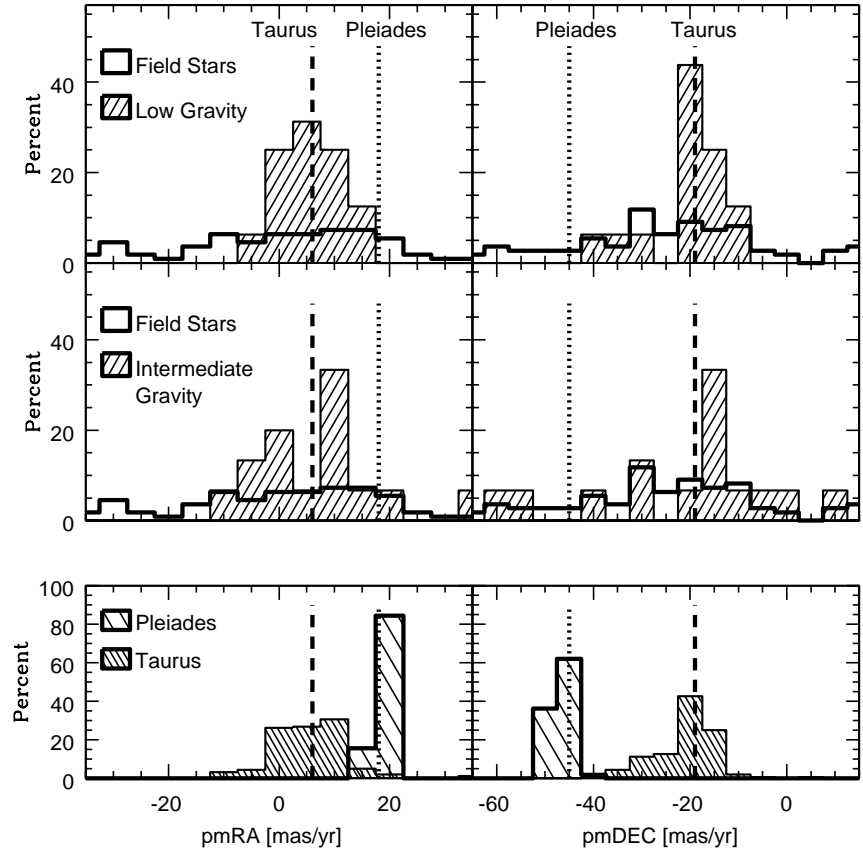}
\caption{Histograms of proper motions extracted from the USNO-B1.0 catalog.  In 
the top and middle panels, open 
histograms represent proper motions for stars which were spectroscopically 
determined to be field dwarfs; hatched histograms
represent proper motions for new young (top) and intermediate-age (middle) PMS 
targets.
In the bottom panels the light hatched histogram shows USNO-B1.0 proper motions
for a sample of Pleiades members; the dark hatched histogram shows equivalent 
data for the sample of known 
Taurus members indicated in Fig.~\ref{fig:spat2}.  
All histograms have been normalized for ease of comparison  
and average values for Taurus and the Pleiades are shown in all panels.
Data have been binned by 5 mas/yr.  The average measured uncertainty for new PMS 
candidates is 4 mas/yr.
}
\label{fig:pm}
\end{figure}

We extracted USNO-B1.0 proper motions\footnote{We define an object to have a measured USNO proper 
motion if it has either a non-zero proper motion
or a non-zero proper motion uncertainty in the USNO-B1.0 Whole Sky catalog.  This cut will 
bias us against objects with intrinsically small
(proper motion $<$2 mas/yr) but well-measured (uncertainty $<$ 2 mas/yr) proper 
motions which will be listed with $\mu$RA=0,
$\mu$Dec=0, 
$\mu$RA\_err=0 and 
$\mu$Dec\_err=0.  However, because the USNO proper motion measurements are rounded to the 
nearest mas/yr and 
non-measurements are not indicated, we are unsure when
a measured proper motion uncertainty of zero refers to a very small uncertainty versus a non-detection.
Therefore, we feel this is a necessary, if 
conservative, selection criteria.} (Monet et al. 2003) for 141 of 
the 190 objects with spectra.
Fig.~\ref{fig:pm} shows histograms of the $\alpha$ and $\delta$ proper motion components  
for new young and intermediate-age PMS 
targets and 
for stars 
classified as field dwarfs.
Because proper motions listed in the USNO catalog are relative rather than 
absolute (i.e., the proper motions of background stars have not been
accounted for), it was necessary to extract similarly derived USNO proper motions of known 
Taurus members for 
comparison rather than using more accurate
values listed in the literature
(e.g., Frink et al. 1997). 
The bottom panels of 
Fig.~\ref{fig:pm} show histograms of USNO proper motions for 
160 known 
Taurus members 
and for 58 Hipparcos-selected Pleiades members
(Robichon et al. 1999).

The proper motions of the PMS stars appear strongly 
correlated (independent of RA) 
and distinct from the proper motions of Pleiades members,
whereas the spectroscopic field dwarfs exhibit a very broad distribution of proper motions.
These results indicate that 
the newly-identified PMS population
is not associated with the Pleiades and is not a collection 
of random field stars.
Further, the PMS stars
farthest from the known Taurus population do not exhibit systematically higher proper motions, as
would have been expected if they had been ejected from the
current star-forming regions, giving further evidence that we have identified a new
region of relatively recent star-formation within the Eastern regions of the clouds.
The young sample in particular appears to have proper motions consistent with 
Taurus.  A possible interpretation of these data is that the new PMS objects
in the central regions are associated with the known Taurus population and
the new PMS objects farther away were formed out of molecular
material sharing a similar velocity and distance to that forming the current 1 Myr-old population.
However, typical USNO proper motion 
uncertainties are large ($\pm$4 mas/yr), and the Taurus data itself exhibit a large 
spread (bottom panels of Fig.~\ref{fig:pm}).
Therefore, a more detailed kinematic study of
these objects is necessary before a definitive conclusion about their origin can be drawn (see Section 5).

\subsection{A New Distributed Population and the Post T-Tauri Star Problem}

There has been much debate over the last few decades as to whether there exists
a population of 3--10 Myr old post T-Tauri stars associated with the current 1--2 Myr old 
Taurus members.
Spectroscopic follow-up studies of RASS-selected sources have identified a wide-spread 
population of stars
in the vicinity of Taurus (e.g., Neuh\"{a}user 
et al. 1997, Wichmann et al. 1996), but the distances, ages, and origins of these stars remain 
controversial.
Measures of x-ray emission and lithium equivalent widths are consistent with any 
ages from $\approx$1--100 Myr for these objects, and
the nearly uniform spatial distribution of this population is consistent with its originating from 
a number of sources other than Taurus (e.g., Brice\~{n}o et al. 1997). 
We have compared the spatial location of the new PMS sources identified here to the distribution of
RASS sources in the area (eg., Guillout et al. 1998) and find no correlation
between areas of dense x-ray sources detected in the RASS 
and the newly-identified stars in the Eastern part
of our survey region.  

Besides X-ray emission, an efficient method of identifying nearby older PMS stars 
over a large area is through
optical and near-infrared imaging combined with spectroscopic follow-up observations.  
Surface gravity signatures present in optical spectra of late-K and M stars
offer the advantage of clearly distinguishing young and intermediate-age
PMS stars from 75--100 Myr-old stars.
Considering the large number of imaging/spectroscopic surveys in the Taurus region, a 
natural question to ask is, how were the PMS stars presented here
not discovered prior to our survey?  

The primary reason these objects were not identified in earlier work is that
we have searched much farther away from the clouds than most previous photometric/spectroscopic 
studies.
The large-area work by Luhman (2006) 
probed only as far East as $\alpha \approx$5 hr and did not extend to the large number of newly
discovered PMS stars in the Eastern portion of our survey at $\alpha>$80 deg. 
Additionally, our survey occupies a unique range in color/magnitude space.  Low mass stars discovered
in early CCD surveys (e.g., Brice\~{n}o et al. 2002) are saturated in the Quest-2 data.  
More recent surveys (e.g., Luhman 2004, Guieu et al. 2006, Luhman 2006) have been concerned primarily 
with finding new brown dwarfs with spectral types $\geq$M6, which will be predominantly fainter and
redder than the candidates discussed here.  
Indeed, comparison of USNO-$I2$ and 2MASS-$J,H,K_S$ photometry  
for the Quest-2 PMS candidates to those selected in the Luhman (2006) 
study reveals that the Quest-2 candidates are systematically bluer.  
Despite the large overlap in area ($\sim$50 deg$^2$), the only observed candidate in our survey 
with spectral type $\geq$M6 selected 
as a candidate via the Luhman (2006) criteria is 
the one
star (SCH J04311908+2335048) which our two works both identified.  
It is unlikely that the intermediate-age objects found in this work,
 which are predominantly bluer
than the young objects (see Fig.~\ref{fig:cmd}), would have been selected as candidate
Taurus members in any of the previous optical studies.

While we have discovered a new distributed population of both young and intermediate-age stars,
we do not at this time claim that these objects represent the post T-Tauri stars in Taurus.
The spatial distribution of the PMS stars identified far from the known members is not
consistent with those stars being associated with the Taurus subclusters.
Additionally, the fact that the proper motions for the new young stars located tens of degrees away
are consistent with known Taurus members
implies that they were not ejected at high velocities from the  current star-forming regions.
Because our dataset is not complete either spatially 
or in magnitude/color space (due to difficulties with weather and calibration), we cannot assess
the full extent of this new population.  Rather, we note that its existence
hints there may be many more as-yet undiscovered PMS objects waiting to be identified in and 
surrounding the 
Taurus clouds.

\section{Summary and Future Work}

We have completed a large-area ($\sim$250 deg$^2$) $B,R,I$ imaging survey of the southern part of 
the 
Taurus molecular cloud using the Quest-2 camera.  From a combination of these 
data and 
2MASS $J,H,K_S$ photometry, candidate PMS objects were selected throughout the 
region.
We have observed spectra of 190 of these objects ($\approx$10\% of the photometrically-selected
candidates) and,
based on comparison to spectra of Taurus members and members of other young associations, have 
determined 42 are bona fide PMS 
stars.  
From the strength of spectral features sensitive to surface gravity,
we subdivide the new PMS objects into young and intermediate-age
stars.
We find members of both populations throughout the survey region and well beyond previously 
studied areas.
In particular we find a new concentration of PMS objects in the Eastern portion 
of the cloud
located at $\alpha \sim$82$^o$ and $\delta \sim 24^o$, which likely did not originate
from the previously known star-forming regions.
We have analyzed the USNO proper motions of 141 spectral targets and find a 
strong concentration
of proper motions for the new PMS stars around those measured for known Taurus 
members.   
Conversely, the spectroscopic field dwarf population exhibits a very broad 
distribution of proper motions.

Radial
velocities from high resolution spectra for all 42 new PMS stars combined with more 
robust proper motion
measurements derived from existing catalogs will 
allow us to
compute a 3D $U,V,W$ velocity for each object over a range of distances.
In addition, lithium equivalent widths can help confirm the youth of the new PMS
sample which is currently based on NaI and KI line strengths.
While lithium depletion occurs slowly over $>$100 Myr for G type stars and is therefore not suitable 
for distinguishing PTTSs from 100 Myr objects at these temperatures,  
for K to mid-M type
stars that are fully convective, lithium depletion occurs over much faster timescales 
($\sim$10 Myr; e.g., D'Antona \& Mazzitelli 1994)
and is a more robust indicator of youth.  
Finally, we emphasize that the new PMS population of 42 stars and a spectroscopic confirmation rate of 
$\sim$20\% implies that several hundred similar young and intermediate-age PMS stars may be present in our larger photometric database.

\section{Acknowledgments}

The authors are appreciative of the Quest-2 collaboration including David Rabinowitz,
Anne Bauer and Jonathan Jerke, for observing and processing the photometric drift scan data.
We would like to thank Ashish Mahabal and Eilat Glikman for many discussions concerning 
the systematics and calibration of the Quest-2 data.  We thank Lee Hartmann for his
insights and suggestions which improved the quality of this manuscript.
We are grateful to assistance from the
entire Palomar staff, in particular Jean Mueller, Karl Dunscombe, and Dipali.   
This manuscript has made use of data from the Two Micron All Sky Survey.

\clearpage

\clearpage
\begin{deluxetable}{cccccccccclrc}
\tabletypesize{\scriptsize}
\tablecolumns{12}
\tablewidth{0pc}
\tablenum{1}
\tablecaption{Measured Quantities for New PMS Stars}
\tablehead{
 \colhead{ID\tablenotemark{a}} & 
 \colhead{b} & 
 \colhead{r} & 
 \colhead{i} & 
 \colhead{J\tablenotemark{c}} &
 \colhead{H\tablenotemark{c}} &
 \colhead{K$_S$\tablenotemark{c}} &
 \colhead{TiO-7140} &
 \colhead{TiO-8165} &
 \colhead{Na-8195} &
 \colhead{SpType} &
 \colhead{$W(H\alpha)$ [\AA]} &
 \colhead{Age}  
}
\startdata
 SCH J0325332+2426581    & 18.0 &  16.4 &  14.7 &  12.34 & 11.72 & 11.47 &  2.02 &  1.30 &  0.89 & M4.5 & -10 & int \\
 SCH J0359099+2009362    & 19.8 &  18.0 &  16.1 &  13.47 & 12.89 & 12.53 &  2.32 &  1.42 &  0.85 & M4.75 & -7  & int \\
 SCH J0400220+2232382    &  --  &  18.4 &  16.3 &  13.45 & 12.76 & 12.45 &  2.23 &  1.36 &  0.87 & M4.75 & -6  & int \\
 SCH J0400279+2031593    & 20.0 &  18.2 &  16.1 &  13.14 & 12.56 & 12.23 &  2.54 &  1.49 &  0.84 &  M5.75 & -16 & int \\
 SCH J0407246+2332554    & 18.4 &  16.7 &  14.8 &  12.77 & 12.06 & 11.85 &  1.69 &  1.16 &  0.88 &    M4 & -6  & int \\
 SCH J0407350+2237396    & 18.6 &  16.8 &  15.0 &  12.16 & 11.60 & 11.25 &  2.38 &  1.51 &  0.86 & M5 & -16 & int \\
 SCH J0412433+2055306    &  --  &  20.3 &  17.8 &  14.24 & 13.53 & 13.17 &  2.97 &  1.95 &  0.90 &    M8 & -15 & int \\
 SCH J0416272+2053093\tablenotemark{b}     & 18.6 &  16.8 &  14.9 &  12.05 & 11.47 & 11.11 &  2.26 &  1.43 &  0.93 &    M5 & -5  & young \\
 SCH J0420068+2432267    & 17.5 &  16.1 &  14.6 &  12.42 & 11.87 & 11.59 &  1.84 &  1.18 &  0.88 &    M4 & --  & int \\
 SCH J0420491+2327370    & 18.3 &  16.7 &  14.9 &  12.07 & 11.39 & 11.09 &  2.03 &  1.25 &  0.88 & M4.25 & -8  &  int \\
 SCH J0426452+2131408    & 19.3 &  17.8 &  15.8 &  13.19 & 12.64 & 12.31 &  2.23 &  1.38 &  0.87 &    M4.75 & -8  & int \\
 SCH J0427074+2215039    &  --  &  17.7 &  15.4 &  12.27 & 11.64 & 11.29 &  2.64 &  1.74 &  0.96 & M6.75 & -18 & young \\
 SCH J0429595+2433080\tablenotemark{b} &  --  &  17.9 &  15.7 &  11.68 & 10.53 &  9.81 &  1.87 &  1.39 &  0.97 &    M5.5 & -71 & young  \\
 SCH J0431191+2335048\tablenotemark{b} &  --  &  20.3 &  17.5 &  13.50 & 12.71 & 12.19 &  2.63 &  2.12 &  0.97 &    M8 & -32 & young \\
 SCH J0433131+2025200    &  --  &  19.1 &  17.0 &  14.20 & 13.46 & 13.14 &  2.36 &  1.40 &  0.87 & M5 & -12 & int \\
 SCH J0434454+2308035    & 20.2 &  18.2 &  16.0 &  12.80 & 12.02 & 11.70 &  2.31 &  1.45 &  0.93 & M5.25 & -11 & young \\
 SCH J0438001+2327167    & 19.7 &  18.0 &  15.9 &  13.27 & 12.66 & 12.34 &  2.41 &  1.45 &  0.85 & M5.25 & -15 & int \\
 SCH J0438163+2326404    & 17.7 &  16.2 &  14.3 &  11.80 & 11.24 & 10.96 &  2.24 &  1.38 &  0.93 &    M4.75 & -7  & young \\
 SCH J0438586+2336352    & 18.4 &  16.8 &  14.3 &  11.97 & 11.36 & 11.03 &  2.03 &  1.19 &  0.91 & M4.25 & -15 & young \\
 SCH J0438587+2323596    & 19.1 &  17.5 &  15.3 &  12.49 & 11.93 & 11.59 &  2.62 &  1.68 &  0.92 & M6.5 & -14 & young \\
 SCH J0439016+2336030    & 16.7 &  15.2 &  12.8 &  11.33 & 10.59 & 10.18 &  2.41 &  1.60 &  0.92 &    M6 & -62 & young \\
 SCH J0439064+2334179    & 19.0 &  17.3 &  14.6 &  12.09 & 11.53 & 11.19 &  2.78 &  1.86 &  0.94 &  M7.5 & -8  & young \\
 SCH J0439410+2304262    &  --  &  19.1 &  17.1 &  14.41 & 13.81 & 13.46 &  2.09 &  1.31 &  0.87 & M4.5 & -12 & int \\
 SCH J0439507+2133564    &  --  &  19.7 &  17.5 &  14.43 & 13.80 & 13.43 &  2.67 &  1.62 &  0.84 & M6 & -8  & int \\
 SCH J0440534+2055473    & 18.8 &  17.1 &  15.1 &  12.48 & 11.88 & 11.62 &  2.32 &  1.44 &  0.87 & M5. & -11 & int \\
 SCH J0502377+2154050    & 19.2 &  17.4 &  15.6 &  13.16 & 12.53 & 12.20 &  1.99 &  1.25 &  0.88 & M4.25 & -10 & int \\
 SCH J0506466+2104298    & 18.1 &  16.5 &  14.6 &  12.05 & 11.40 & 11.11 &  2.43 &  1.51 &  0.91 & M5.25 & -14 & young \\
 SCH J0516021+2214530    & 17.4 &  15.9 &  14.1 &  11.67 & 11.13 & 10.75 &  2.25 &  1.39 &  0.90 &    M5 & -10 & young \\
 SCH J0516058+2236152    & 19.1 &  17.2 &  15.6 &  13.29 & 12.59 & 12.30 &  1.88 &  1.18 &  0.88 &    M4 & -8  & int \\
 SCH J0518028+2327126    & 18.6 &  17.2 &  15.3 &  12.99 & 12.32 & 11.88 &  2.03 &  1.47 &  0.88 & M5 & -21 & young \\
 SCH J0522333+2439254    & 18.9 &  17.0 &  15.2 &  12.75 & 12.04 & 11.72 &  2.18 &  1.38 &  0.90 &    M4.75 & -7  & young \\
 SCH J0522335+2439197    & 19.2 &  17.2 &  15.3 &  12.79 & 12.14 & 11.79 &  2.10 &  1.30 &  0.91 & M4.5 & -8  & young \\
 SCH J0523020+2428087    & 18.1 &  16.4 &  14.8 &  12.60 & 11.92 & 11.61 &  1.78 &  1.16 &  0.90 &    M4 & -3  & int \\
 SCH J0523500+2435237    &  --  &  18.7 &  16.6 &  13.81 & 13.14 & 12.77 &  2.63 &  1.61 &  0.88 & M6 & -18 & young \\
 SCH J0531021+2333579    & 17.5 &  16.0 &  14.0 &  12.28 & 11.64 & 11.39 &  1.79 &  1.18 &  0.90 &    M4 & -5  & int \\
 SCH J0531026+2334022    & 17.6 &  16.0 &  13.9 &  12.26 & 11.58 & 11.35 &  1.83 &  1.22 &  0.89 &    M4 & -5  & int \\
 SCH J0532021+2423030    & 19.9 &  18.2 &  16.2 &  13.69 & 13.06 & 12.80 &  2.50 &  1.44 &  0.85 & M5 & -19 & int \\
 SCH J0533363+2102276    & 17.4 &  15.8 &  14.2 &  11.93 & 11.32 & 11.07 &  2.05 &  1.28 &  0.88 &  M4.5 & -7  & int \\
 SCH J0534480+2243142    & 18.2 &  16.6 &  15.0 &  12.83 & 12.18 & 11.93 &  1.94 &  1.22 &  0.89 & M4.25 & -6  & young \\
 SCH J0536190+2242428    & 17.7 &  16.3 &  14.5 &  12.13 & 11.53 & 11.27 &  2.17 &  1.35 &  0.89 & M4.75 & -12 & int \\
 SCH J0537385+2428518    & 17.8 &  16.2 &  14.2 &  11.66 & 11.06 & 10.78 &  2.44 &  1.49 &  0.89 & M5.25 & -14 & young \\
 SCH J0539009+2322081    & 19.1 &  17.6 &  15.3 &  12.68 & 12.10 & 11.79 &  2.53 &  1.59 &  0.88 &    M6 & -16 & young \\
\enddata
\tablenotetext{a}{Object IDs given in J2000 coordinates.}
\tablenotetext{b}{Three of our PMS objects were previously identified in the literature:
SCH J0429595+2433080 (Gueiu et al. 2006), SCH J0431191+2335048 (Luhman 2006), and SCH J0416272+2053093 (Wichmann et al. 1996)}
\tablenotetext{c}{Near-infrared photometry taken from 2MASS.}

\end{deluxetable}

\end{document}